\documentclass[twocolumn,showpacs,superscriptaddress,longbibliography,prl]{revtex4-1}

\usepackage{graphicx}
\usepackage{dcolumn}
\usepackage{amsmath}
\usepackage{color}
\usepackage{braket}
\usepackage{float}
\usepackage[english]{babel}

\newcommand{\BaCoVO}{BaCo$_2$V$_2$O$_8$}

\begin{document}

\title{Quantum Criticality of an Ising-like Spin-1/2 Antiferromagnetic Chain in Transverse Magnetic Field}

\author{Zhe~Wang}
\email{zhe.wang@hzdr.de}
\affiliation{Experimental Physics V, Center for Electronic
Correlations and Magnetism, Institute of Physics, University of Augsburg, 86135 Augsburg, Germany}
\affiliation{Institute of Radiation Physics, Helmholtz-Zentrum Dresden-Rossendorf, 01328 Dresden, Germany}

\author{T. Lorenz}
\email{tl@ph2.uni-koeln.de}
\affiliation{Institute of Physics II, University of Cologne, 50937 Cologne, Germany}

\author{D.~I.~Gorbunov}
\author{P.~T.~Cong}
\affiliation{Dresden High Magnetic Field Laboratory (HLD-EMFL), Helmholtz-Zentrum
Dresden-Rossendorf, 01328 Dresden, Germany}

\author{Y. Kohama}
\affiliation{Institute for Solid State Physics, University of Tokyo, Kashiwa, Chiba 277-8581, Japan}

\author{S. Niesen}
\author{O. Breunig}
\author{J. Engelmayer}
\author{A. Herman}
\affiliation{Institute of Physics II, University of Cologne, 50937 Cologne, Germany}

\author{Jianda~Wu}
\email{jdwu@pks.mpg.de}
\affiliation{Max-Planck-Institut f\"{u}r Physik komplexer Systeme, 01187 Dresden, Germany}

\author{K. Kindo}
\affiliation{Institute for Solid State Physics, University of Tokyo, Kashiwa, Chiba 277-8581, Japan}

\author{J.~Wosnitza}
\author{S.~Zherlitsyn}
\affiliation{Dresden High Magnetic Field Laboratory (HLD-EMFL), Helmholtz-Zentrum
Dresden-Rossendorf, 01328 Dresden, Germany}

\author{A.~Loidl}
\affiliation{Experimental Physics V, Center for Electronic
Correlations and Magnetism, Institute of Physics, University of Augsburg, 86135 Augsburg, Germany}

\date{\today}

\begin{abstract}
We report on magnetization, sound velocity, and magnetocaloric-effect measurements of the Ising-like spin-1/2 antiferromagnetic chain system BaCo$_2$V$_2$O$_8$ as a function of temperature down to 1.3~K and applied transverse magnetic field up to 60~T.
While across the N\'{e}el temperature of $T_N\sim5$~K anomalies in magnetization and sound velocity confirm the antiferromagnetic ordering transition,
at the lowest temperature the field-dependent measurements reveal a sharp softening of sound velocity $v(B)$ and a clear minimum of temperature $T(B)$ at $B^{c,3D}_\perp=21.4$~T, indicating the suppression of the antiferromagnetic order.
At higher fields, the $T(B)$ curve shows a broad minimum at $B^c_\perp = 40$~T,
accompanied by a broad minimum in the sound velocity and a saturation-like magnetization.
These features signal a quantum phase transition which is further characterized by the divergent behavior of the Gr\"{u}neisen parameter $\Gamma_B \propto (B-B^{c}_\perp)^{-1}$.
By contrast, around the critical field, the Gr\"{u}neisen parameter converges as temperature decreases,
pointing to a quantum critical point of the one-dimensional transverse-field Ising model.
\end{abstract}

\maketitle

Phase transitions between distinct phases of matter can take place
even at zero temperature due to quantum fluctuations \cite{Sachdev99}.
Typical quantum phase transitions are induced
by tuning external parameters, such as magnetic field, pressure, or chemical doping.
Understanding the quantum phase transitions has become one of the most significant topics in condensed-matter physics \cite{Sachdev08,Vojta03,Lnsen07,Gegenwart08,Giamarchi08,SachdevKeimer11}.
It is generally believed that universal scaling occurs near the quantum critical point (QCP),
which can be thermodynamically characterized by a divergent Gr\"{u}neisen parameter \cite{Zhu03,Garst05,Wu11},
as experimentally found in heavy-fermion compounds \cite{Lnsen07,Gegenwart08} as well as in the one-dimensional (1D) Heisenberg spin systems \cite{Wolf11,Lorenz08,Breunig17}.
However, very recently it is theoretically shown that for the QCP of the transverse-field Ising (TFI) spin chain,
the Gr\"{u}neisen parameter does not diverge when the QCP is approached by decreasing temperature at the critical field \cite{Wu18},
which appeals for further experimental efforts on the study of quantum critical behavior of the paradigmatic TFI chain.

\begin{figure}[t]
\centering
\includegraphics[width=80mm,clip]{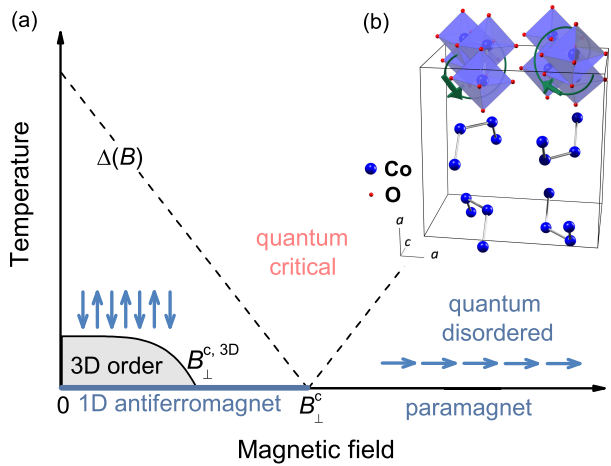}
\vspace{2mm} \caption[]{\label{Fig:PD_TFIC}
(a) At zero temperature, an isolated Ising chain exhibits a spin-gapped antiferromagnetic long-range order,
before the spin gap $\Delta (B)$ is closed at the transverse critical field $B^c_\perp$.
Above $B^c_\perp$, a quantum disordered paramagnetic phase is induced.
Weak inter-chain couplings can stabilize a 3D order at finite temperature and in fields up to $B^{c,3D}_\perp$.
(b) Unit cell of BaCo$_2$V$_2$O$_8$ with screw chains constituted by the edge-shared CoO$_6$ octahedra \cite{He05,He06,Niesen13}.
Ba and V ions are omitted for clarity.
}
\end{figure}

The TFI chain plays an important role in quantum statistical and condensed-matter physics \cite{Sachdev99,Suzuki13,Dutta15},
because quantitative understanding of the relevant physics can be achieved in an exact sense, based on its rigorous solvability by analytical as well as numerical methods \cite{McCoy68,Lieb61,Pfeuty70,Zamolodchikov89,Orus2008,Wu14,Calabrese11,Heyl15,Caux08}.
Without magnetic field, the ground state of an isolated Ising spin chain with nearest-neighbour antiferromagnetic exchange interactions
corresponds to a spin-gapped long-range antiferromagnetic order [Fig.~\ref{Fig:PD_TFIC}(a)].
By applying an external transverse field, the spin gap is reduced and finally closed at the critical field $B^c_\perp$,
which leads to a quantum-disordered and gapped phase at higher fields [Fig.~\ref{Fig:PD_TFIC}(a)].
At the corresponding QCP, the spatial spin-correlation functions decay in power law, as established for the universality class of the 1D TFI model \cite{McCoy68,Pfeuty70}.
Further studies have shown that even in presence of perturbative XY-exchange interactions (i.e., the Ising-like XXZ model),
the spin excitations remain gapped at zero field \cite{Shiba80,Bougourzi98}, and the spin correlations decay in the same way at the transverse-field-induced quantum phase transition, suggesting that the associated QCP belongs to the same universality class of the 1D TFI model \cite{Dmitriev02}.

While theoretical efforts have provided detailed understanding of the quantum phase transition of the 1D TFI model \cite{Sachdev99,Vojta03,Suzuki13,Calabrese11,Heyl15,Pfeuty70,Zamolodchikov89,Dmitriev02,Dutta15,Wu14,Wu18,Mahdavifar17,Cole17},
it remains very challenging to experimentally realize this paradigmatic model in a real material for the study of quantum critical behavior  \cite{Bitko96,Coldea10,Imai14}.
This is because several key conditions have to be fulfilled simultaneously:
strong easy-axis anisotropy is required;
spin-spin interaction should be dominated by intra-chain coupling;
and the intra-chain coupling should not be too large, so that the critical field is experimentally accessible.
Here, by performing magnetization, sound velocity, and magnetocaloric-effect measurements as a function of temperature and transverse magnetic field in the Ising-like spin-1/2 antiferromagnetic chain material BaCo$_2$V$_2$O$_8$,
we identify a 1D quantum critical regime around 40 T where long-range magnetic order is suppressed by the transverse field and the thermal energy is much smaller than the dominant intra-chain exchange interaction.
In this regime, our experimental results show that the magnetic Gr\"{u}neisen parameter follows a universal divergence
on approaching the quantum critical point as a function of the magnetic field, but it converges as a function of decreasing temperature.
These features jointly point to the 1D TFI universality class of the underlying QCP \cite{Wu18}.

BaCo$_2$V$_2$O$_8$ hosts screw chains of edge-sharing CoO$_6$ octahedra with the screw axis parallel to the \emph{c} direction of its tetragonal structure [Fig.~\ref{Fig:PD_TFIC}(b)] \cite{He05,He06,Niesen13}.
Due to the crystal-field effects and spin-orbit coupling \cite{Shiba03,Lines63},
the exchange interactions between the Co$^{2+}$ ions can be described
by an effective spin-1/2 antiferromagnetic chain model (the 1D XXZ model) with the nearest-neighbor exchange interaction of $J\sim5$~meV ($\simeq 60$~K) and a pronounced Ising anisotropy with the magnetic easy axis along \emph{c}, as is reflected
by the anisotropic \emph{g} factors $g_\parallel/g_\perp\sim2$ \cite{He06,Grenier15,Kimura2013}.
Even in high magnetic fields, this effective model provides a valid description, since the Zeeman interaction is sufficiently small compared with the spin-orbit coupling \cite{Kimura2013,Wang16,Wang18}.
The local Ising axes are slightly tilted from the \emph{c} axis reflecting the fourfold screw-axis symmetry, which leads to an additional but weaker in-plane anisotropy \cite{Niesen13,Kimura2013,Faure17,Footnote1}.
In zero field, a 3D long-range antiferromagnetic order is stabilized at $T_N\sim$~5~K
by weak inter-chain couplings \cite{Niesen13,He05}.
Our present study reveals that the 3D order is overcome by a transverse field of $B^{c,3D}_\perp = 21.4$~T along the crystallographic [110] direction, which is much smaller than the corresponding 1D critical field $B^c_\perp = 40$~T. Owing to the large difference between $B^{c,3D}_\perp$  and  $B^c_\perp$, BaCo$_2$V$_2$O$_8$ opens the way to explore the 1D quantum criticality.

High-quality single crystals of \BaCoVO~were grown using the floating-zone method \cite{Niesen13}.
For ultrasound and magnetocaloric-effect experiments
a flat sample with a [110] surface of about $3 \times 3$~mm$^2$ and a thickness of 1~mm was used, while the magnetization was measured on a bar-shaped sample of 2~mm along $[110]$ and a cross section of about 1~mm$^2$.
Magnetic fields were applied along $[110]$ for all the measurements.
Magneto-elastic properties were investigated by
measuring the velocity and attenuation of sound waves with
wave vector $k \parallel B$  for longitudinal and transverse polarization  $u \parallel k$  and $u \perp k$, respectively.
The pulsed fields for the ultrasound measurements had a rise time of 33~ms and a pulse duration of 150~ms, and of 7~ms and 20~ms, respectively, for the magnetization measurements.
Magnetocaloric-effect measurements were performed under quasi-adiabatic conditions. The sample was kept thermally isolated in high vacuum during field pulses with a rise time of 14~ms and duration of 36~ms \cite{Kihara13}.

\begin{figure}[t]
\centering
\includegraphics[width=70mm,clip]{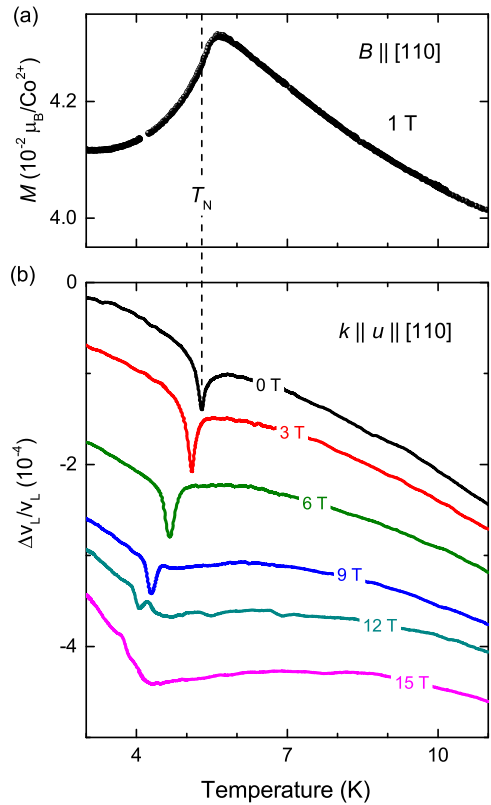}
\vspace{2mm} \caption[]{\label{Fig:TdepTN}
(a) Magnetization as a function of temperature, measured with 1~T field along the [110] direction.
(b) Temperature dependence of the sound-velocity change $\Delta v(T)/v$ for the longitudinal model at magnetic fields up to 15~T.
}
\end{figure}

Figure~\ref{Fig:TdepTN}(a) shows the magnetization for a transverse field of 1~T as a function of temperature.
With decreasing temperature, the magnetization increases and exhibits a sudden drop with a maximum slope at $T_N$, signaling the phase transition to the antiferromagnetic phase \cite{Niesen13}.
At the phase transition, the sound velocity $\Delta v(T)/v$ in zero field exhibits a sharp softening [Fig.~\ref{Fig:TdepTN}(b)].
The strong magnetoelastic coupling in BaCo$_2$V$_2$O$_8$ makes the ultrasound measurement a sensitive probe for detecting the magnetic phase transitions \cite{Yamaguchi2011}.
With increasing field up to 15~T, the ordering phase boundary shifts moderately to lower temperatures, as reflected by the $\Delta v(T)/v$ curves [Fig.~\ref{Fig:TdepTN}(b)], in agreement with the thermal-expansion measurements \cite{Niesen13}.

\begin{figure}[t]
\centering
\includegraphics[width=75mm,clip]{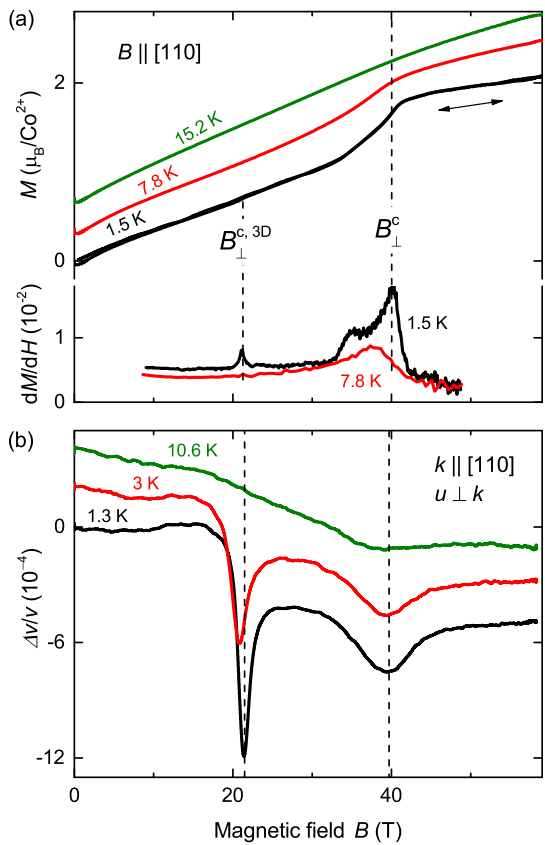}
\vspace{2mm} \caption[]{\label{Fig:MandVeloHF}
(a) Magnetization $M$, its field derivative $dM/dH$, and (b) sound velocity $\Delta v(B)/v$ as a function of magnetic field $B \parallel [110]$ at various temperatures.
At the lowest temperatures, both magnetization and sound velocity exhibit two anomalies at $B^{c,3D}_\perp=21.4$~T and $B^c_\perp=40$~T, respectively, as indicated by dashed lines.
Above $T_N$, the anomalies at $B^{c,3D}_\perp$ disappear, while the anomalies at $B^c_\perp$ systematically broaden.
The curves of different fields are shifted vertically for clarity.
}
\end{figure}

\begin{figure}[t]
\centering
\includegraphics[width=73mm,clip]{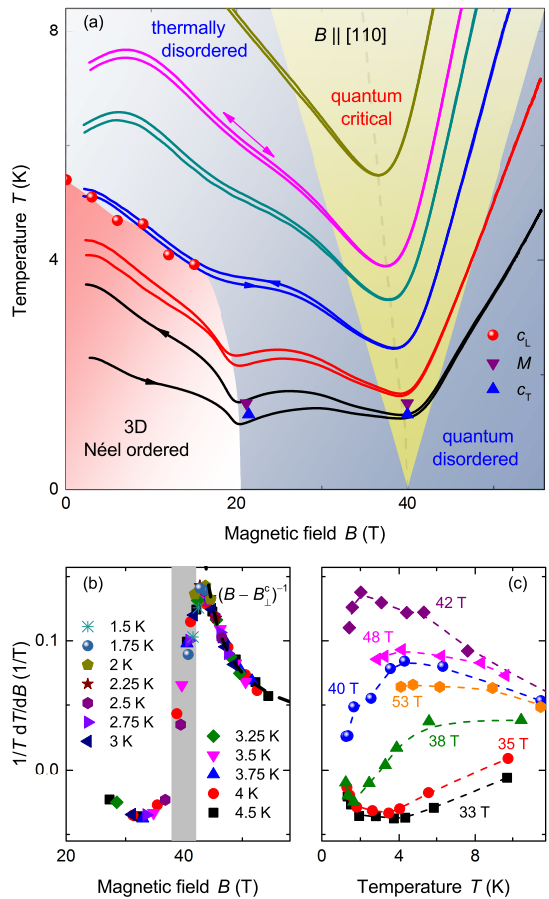}
\vspace{2mm} \caption[]{\label{Fig:PFieldMCE_H110}
(a) Magnetocaloric-effect measurements with various starting temperatures.
The $T(B)$ curves below $T_N$ exhibit clear minima at the 3D phase boundary and at the 1D quantum phase transition.
The latter are present also at temperatures well above $T_N$, and a linear extrapolation (dashed line) marks the QCP at $B^c_\perp = 40$~T.
Anomalies observed in the magnetization $M$ [Fig.~\ref{Fig:MandVeloHF}(a)] and sound velocity curves for $c_{L}$ [Fig.~\ref{Fig:TdepTN}(b)] and $c_{T}$ [Fig.~\ref{Fig:MandVeloHF}(b)] are marked.
For the pulse-field measurements, only the lowest-temperature results are shown because of the large magnetocaloric effects at higher temperatures.
The boundary of the quantum critical regime at lower fields is given by a linear interpolation of the zero-field gap to zero at 40~T, and this regime is symmetrized with respect to the critical field.
In (b) and (c) the experimental values of the Gr\"{u}neisen parameter $\frac{1}{T}\frac{dT}{dB}$ are shown as a function of magnetic field and temperature, respectively.
In (b), above $B^c_\perp$, the data at different temperatures follow a diverging behavior $\Gamma_B\sim(B-B^{c}_\perp)^{-1}$ (dashed line) \cite{Garst05,Wu11}.
In contrast, for the fields around $B^c_\perp$ shown in (c), $\Gamma_B(T)$ clearly converges towards lowest temperatures.
Dashed lines in (c) are guides for the eyes.
}
\end{figure}

Figure~\ref{Fig:MandVeloHF} shows the magnetization and sound velocity as a function of magnetic field up to 60~T for various temperatures.
At 1.3~K, the sound velocity exhibits a very sharp minimum at $B^{c,3D}_\perp=21.4$~T, which shifts to lower field with increasing temperature and vanishes above $T_N$ [Fig.~\ref{Fig:MandVeloHF}(b)].
As the lower-field measurements reveal a moderate decrease of $T_N$, i.e. $dT_N/dB \simeq -0.1$ K/T [Fig.~\ref{Fig:TdepTN}(b)], the sharp minimum in $\Delta v(B)/v$ strongly evidences a suppression of the 3D order at $B^{c,3D}_\perp$.
In the field derivative of the magnetization [Fig.~\ref{Fig:MandVeloHF}(a)],
a weak anomaly appears below $T_N$ at the same field $B^{c,3D}_\perp$, that above $T_N$ disappears, confirming the suppression of the 3D order.
As will be discussed in the following, the field-induced phase transition is further confirmed by magnetocaloric-effect measurements [Fig.~\ref{Fig:PFieldMCE_H110}(a)].

With further increasing field, the sound velocity exhibits a broader minimum at $B^c_\perp=40$~T and levels off above 50~T.
This minimum is evidently observable even above $T_N$ at slightly lower fields.
Very similar behavior is revealed by the magnetization measurements.
At 1.5~K, the magnetization increases continuously with increasing magnetic field and
a saturation-like feature is observed above $B^c_\perp$ \cite{Kimura2013}, which is typical for the TFI chain \cite{Suzuki13}.
This feature persists at higher temperatures above $T_N$ although becoming less pronounced.

In order to further characterize the quantum critical behavior,
we measured the magnetocaloric effects in fields up to 55~T.
The temperature changes $T(B)$ are shown in Fig.~\ref{Fig:PFieldMCE_H110}(a) for various starting temperatures in zero field.
Starting from 2.2~K in the 3D ordered phase, $T(B)$ decreases with increasing $B$ and exhibits a minimum of 1.2~K at 20~T,
indicating the field-induced suppression of the 3D antiferromagetic order.
Upon further increase of the field, the temperature exhibits another broader minimum around $B^c_\perp=40$~T,
which is followed by a continuous increase of $T(B)$ up to the highest fields.
During the down-sweep of the field, the sample temperature essentially follows the up-sweep curve and shows again a minimum at $B^c_\perp$.
The hysteresis below $B^c_\perp$ can be due to field-driven reorientation of twin-domains \cite{Niesen13}, other slow dynamical processes or imperfect adiabatic condition \cite{Zapf14,Nomura16}.

The field-dependent $T(B)$ essentially reflects the inverse field dependence of the magnetic entropy,
because under adiabatic conditions an entropy increase (decrease) of the spin degrees of freedom will absorb (release) heat from (to) the lattice degrees of freedom and result in a decrease (increase) of temperature \cite{Zhu03,Garst05,Wu11,Kihara13}.
Adiabatically magnetizing a paramagnet, for example, causes a temperature increase, as observed above $T_N$ in the low-field range of the $T(B)$ curves.
For the spin-gapped system BaCo$_2$V$_2$O$_8$, where the gap decreases in a transverse field,
the initial temperature increase will change over to a strong decrease with further increasing field,
since the entropy of the spin degrees of freedom significantly increases with the decrease of the spin gap.
Finally, the $T(B)$ curve reaches its minimum at the 1D critical field where the gap is closed,
and above the critical field the temperature increases again,
because another spin gap opens and increases with increasing field which leads to a decreasing spin entropy.
The minimum in $T(B)$ around 40~T reflects the maximum accumulation of entropy above the 1D QCP in BaCo$_2$V$_2$O$_8$, whereas the minimum around 20~T indicates the entropy accumulation at the phase boundary of the 3D order, which results in the broad temperature maximum between 20 and 40~T at the lowest temperatures.

A characteristic signature of a quantum phase transition is the divergence of the magnetic Gr\"{u}neisen parameter $\Gamma_B$, which is defined as the ratio of the temperature derivative of the magnetization to the magnetic specific heat $C_\mathrm{mag}$, i.e., $\Gamma_B = -\frac{dM/dT}{C_\mathrm{mag}}=\frac{1}{T}\frac{dT}{dB}$, under adiabatic conditions \cite{Zhu03,Garst05,Wu11}.
At our experimental temperatures, the phononic specific heat is much smaller than the magnetic contribution (see Ref.~\cite{He06}) and is magnetic-field independent, while contributions of nuclear spins to the specific heat become relevant at much lower temperatures.
Therefore, in the concerned quantum critical regime, the total specific heat $C_\mathrm{total}$ and in particular its magnetic-field dependence is dominated by the electronic spin degrees of freedom, and the observed minima in the $T(B)$ curves are naturally traced back to the non-monotonic magnetic-field dependence of the spin entropy.
Hence, we can approximate the magnetic Gr\"{u}neisen ratio by $-\frac{dM/dT}{C_\mathrm{total}}= \frac{1}{T}\frac{dT}{dB}$,
as obtained experimentally from the magnetocaloric-effect measurements.

As shown in Fig.~\ref{Fig:PFieldMCE_H110}(b), the measured $\frac{1}{T}\frac{dT}{dB}$ values
collapse on a single curve with a sign change close to $B^{c}_\perp$, which is a strong indication of quantum criticality.
The relatively rich experimental data set for $B>B^{c}_\perp$ allows a fit to the critical divergence
$G_r(B-B^{c}_\perp)^{-1}$ with $B^{c}_\perp = 40\pm2$~T and the corresponding prefactor $G_r=0.5\pm0.3$ [dashed line in Fig.~\ref{Fig:PFieldMCE_H110}(b)].
Such a divergent behavior is expected above the field induced QCPs of both the Heisenberg and the Ising spin chain models with $G_r=1$ for $T=0$, but close to and below $B^{c}_\perp$ strong differences are expected for the two models \cite{Garst05,Wu18,Wolf11,Breunig17}.
At low temperatures, the Gr\"{u}neisen ratio of the ideal TFI chain is almost antisymmetric with respect to $B^{c}_\perp$, and becomes exactly antisymmetric at $T=0$.
As $\Gamma_B(T>0, B=B^{c}_\perp)$ is constant for the ideal TFI chain \cite{Wu18},
at low temperatures the sign change of $\Gamma_B$ occurs very close to $B^{c}_\perp$.
In contrast, in the Heisenberg spin chain the Gr\"{u}neisen ratio is asymmetric with respect to $B_c$, and diverges as $1/T$ at $B_c$, which results in a significant shift of the sign-change position away from $B_c$ \cite{Breunig17}.
As shown in Fig.~\ref{Fig:PFieldMCE_H110}(c), a divergence of $\Gamma_B(T,B = B^{c}_\perp)$ with decreasing temperature is clearly absent.
The data reveal a converging Gr\"{u}neisen parameter at $B^{c}_\perp$, which is a characteristic feature of the TFI-chain QCP \cite{Wu18}.

To conclude, by performing magnetization, sound velocity, and magnetocaloric-effect measurements, our study provides thermodynamic evidence for the existence of a 1D quantum phase transition in BaCo$_2$V$_2$O$_8$ at 40~T, as well as for a field-induced suppression of 3D order at 21.4~T.
Our experimental results reveal that while at the lowest temperature the Gr\"{u}neisen parameter follows a universal divergence towards the critical field, in the quantum critical regime it converges with decreasing temperature around the critical field,
pointing to a quantum phase transition of the 1D transverse-field Ising model.

\begin{acknowledgements}
We would like to thank Qimiao Si for fruitful discussions. We acknowledge partial support by the Deutsche Forschungsgemeinschaft via the Transregional Research Collaboration TRR 80: From Electronic Correlations to Functionality (Augsburg - Munich - Stuttgart), the Collaborative Research Center CRC 1238: Control and Dynamics of Quantum Materials (Cologne, Projects A02 \& B01), and the Collaborative Research Center SFB 1143: Correlated Magnetism: From Frustration to Topology (Dresden). The high-field experiments at Dresden were supported by HLD at HZDR, member of the European Magnetic Field Laboratory (EMFL).
\end{acknowledgements}

\end{document}